\documentclass[twocolumn,showpacs,preprintnumbers,amsmath,amssymb,aps,prl]{revtex4}
\usepackage{graphicx}
\begin{document}

\title{
Viscous Decoupling Transitions For Individually  Dragged Particles 
in Systems with Quenched Disorder}   
\author{    
 C. J. Olson Reichhardt and C. Reichhardt} 
\affiliation{
 Theoretical Division,
Los Alamos National Laboratory, Los Alamos, New Mexico 87545 
} 

\date{\today}
\begin{abstract}
We show that when an individual particle is dragged through 
an assembly of other particles in the presence of quenched disorder, 
a viscous decoupling transition occurs between the dragged particle 
and the surrounding particles which is controlled by
the quenched disorder. 
A counterintuitive consequence of this transition is that the velocity of the
dragged particle can be increased
by increasing
the strength or density of the quenched disorder.
The decoupling transition can also occur 
when
the external drive on the 
dragged particle is increased, and is
observable as a clear signature in the velocity-force response.
\end{abstract}
\pacs{82.70.Dd,74.25.Qt}
\maketitle

\vskip2pc
Recently, several experiments have demonstrated the effectiveness
of a local probing technique in which an
individual particle is dragged through an assembly of other particles
\cite{Weeks,Meyer,Cara,Bechinger}.  The velocity-force
response of the driven particle was measured for a single magnetic
colloid pulled through a background of nonmagnetic colloids in 
Ref.~\cite{Weeks}, while in other experiments, optical tweezers were used
to drive a single colloid through a collection of charge stabilized
colloids \cite{Cara,Bechinger}.
Several numerical works have predicted that a rich variety of dynamical
regimes can occur in these types of systems   
depending on whether the probe particle is driven through a 
glassy media \cite{Hastings,Olson,Brady,Evans,Chandler},
a crystal \cite{Reichhardt}, or an assembly of rods \cite{Lowen}.  
The colloidal dragging experiments are performed on the microscale;       
however, it is now becoming possible to conduct similar 
experiments on the meso- and nanoscales. 
For example, 
recent 
experiments have shown that individual vortices in
type-II superconductors
can be manipulated using a magnetic force microscope tip
\cite{Moler1,Moler2}.
This technique could be used to study vortex
entanglement \cite{Ent} and the depinning of vortices from extended
defects \cite{Nelson}. 
A simpler initial study would be to 
measure the drag forces on a single driven 
vortex in the presence of other vortices and quenched disorder from the
sample. 
Currently,
no predictions exist for the response in this regime.
 
In previous studies of single colloids driven through colloidal assemblies,
the driven or probe colloid interacted only with the surrounding colloids.
In the superconducting vortex system or in colloidal systems containing
random quenched disorder, both the probe particle and the surrounding particles
interact with the underlying quenched disorder which acts
as a pinning potential. 
The addition of pinning might be expected merely
to 
increase the overall drag on the probe particle; however, in this work 
we show that pinning can induce 
counterintuitive 
changes in the probe particle motion 
since the pinning couples to {\it both} the
probe and the surrounding particles. 
For example, 
increasing
the strength or density of pinning sites can induce a 
decoupling transition which 
sharply {\it reduces} 
the 
effective drag on the probe particle. 
The same decoupling transition 
occurs 
when the driving force on the probe particle is increased,
and 
appears
as a clear jump in the velocity-force curves. 
In the strongly damped regime, the
probe particle couples strongly to the surrounding particles and 
induces
irreversible plastic deformations. 
When the pinning strength is increased, the surrounding particles 
are trapped by pins
and cannot respond 
to the probe particle, so
the probe particle
decouples from the surrounding particles 
and the effective damping is reduced.
Similarly, when the drive is sufficiently 
large, the 
probe particle is unable to induce topological 
rearrangements in the surrounding particles 
and a decoupling transition occurs.
We specifically examine 
a system with screened Coulomb particle-particle interactions.
Experimentally, this corresponds to a two-dimensional colloidal system 
with quenched disorder, such as 
in Ref.~\cite{Ling}. 
We also 
study interactions appropriate for 
vortices in type-II superconductors.    

We consider $N$ particles in a two-dimensional sample  
of size $L \times L$
with periodic boundary conditions in the
$x$ and $y$ directions. 
The overdamped equation of motion for particle $i$ is  
\begin{equation}
\eta \frac{d{\bf r}_i}{dt} = {\bf F}^i_{s} + {\bf F}^i_{p} 
+ {\bf F}^i_{ext} + {\bf F}_i^{T}  
\end{equation} 
where $\eta$ is the damping constant which we set to $\eta=1$.
The force from particle-particle interactions is 
${\bf F}^i_{s}=-F_{0}\sum_{i\not=j}^{N}\nabla V(r_{ij})$,
where
$F_{0} = Z^{*2}/(4\pi\epsilon \epsilon_{0})$, $Z^{*}$ is the unit of charge, 
$\epsilon$ is the dielectric constant of the medium, and the distance 
between particles located
at ${\bf r}_i$ and ${\bf r}_j$ is
$r_{ij} = |{\bf r}_{i} - {\bf r}_{j}|$.
We use a Yukawa potential, $V(r_{ij}) =  \exp(-\kappa r_{ij})/r_{ij}$, 
with screening length $1/\kappa=2$.
The quenched disorder is modeled as
$N_p$ randomly placed parabolic pinning traps with pinning force
${\bf F}^i_p=\sum_k^{N_p}f_p(r_{ik}/r_p)\Theta(r_p-r_{ik}){\hat {\bf r}}_{ik}$,
where $f_p$ is the pinning strength, $r_p=0.25$ is the pinning radius, 
$\Theta$ is the Heaviside step function, $r_{ik}=|{\bf r}_i-{\bf r}^p_k|$,
${\bf r}_k^p$ is the location of pin $k$, and 
${\hat {\bf r}}_{ik}=({\bf r}_i-{\bf r}^p_k)/r_{ik}$. 
The pinning density is $n_{p} = N_{p}/L^2$ and the particle
density is $n=N/L^2$.
We have examined several different system sizes
and present results for $L = 48$. 
For the probe particle, the external driving force 
${\bf F}^i_{ext}=F_{ext}{\bf \hat x}$, and for all remaining particles,
${\bf F}^i_{ext}=0$.
We construct a velocity-force curve by measuring the 
time-averaged velocity of the probe 
particle, $v=\langle {\bf v}\cdot {\bf{\hat x}}\rangle$, 
as a function of $F_{ext}$.
The thermal force ${\bf F}^{T}_i$ arises from random Langevin kicks 
with the properties $\langle {\bf F}^{T}_{i}\rangle = 0$ and 
$\langle{\bf F}^{T}_{i}(t){\bf F}^{T}_{j}(t^{\prime})\rangle = 2\eta k_{B}T\delta(t - t^{\prime})\delta_{ij}$.   
Unless otherwise noted, we set $T=0$.
We also consider the case of vortices in type-II superconductors, 
modeled as in Ref.~\cite{Periodic} 
where 
${\bf F}^{i}_{s} = \sum^{N}_{i\neq j}f_{0}K_{1}(r_{ij}/\lambda){\hat {\bf r}}_{ij}$.  
Here $K_{1}$ is a modified Bessel function, 
$f_{0} = \phi_{0}/(2\pi\mu_{0}\lambda^3)$,
$\phi = h/2e$ is the flux quantum, 
and 
$\lambda$ is the London penetration depth.
The probe particle in our system recrosses the same pinned region repeatedly
during the course of a measurement; to verify that this does not affect the
results, we tested driving forces applied at low angles to the $x$-axis
which cause the probe to cross a different portion of the pinned region with
each pass through the sample.  We find the same results for all driving
directions.

\begin{figure}
\includegraphics[width=3.5in]{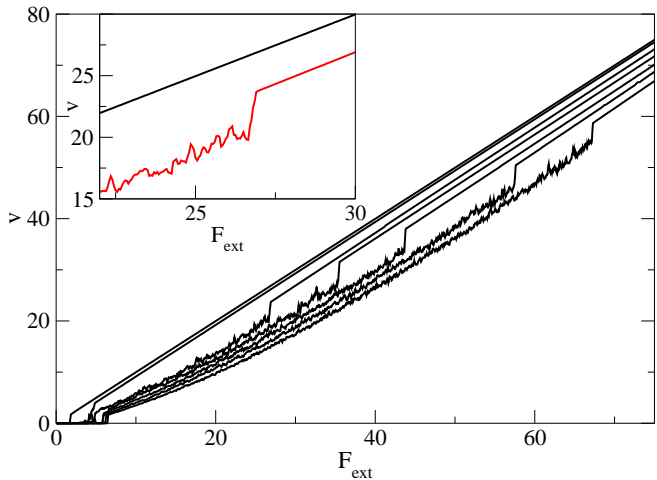}
\caption{
The velocity-force ($v-F$) curves for a single driven probe particle 
in a system
with $n_{p} = 0.85$ and $f_{p} = 2.0$ at        
particle densities $n = 0$, 0.181, 0.292, 0.375, 0.484, 0.53,
and $0.59$ (from top to bottom). 
Inset: 
The transition region for $n = 0.375$ (bottom) 
along with the $n = 0.181$ curve (top).   
}
\end{figure}

We first study a system of Yukawa interacting particles
and examine velocity force ($v-F$) relations for the probe particle 
in the absence of thermal fluctuations for varying particle densities $n$. 
Figure 1 shows the $v-F$ curves  
for fixed 
$n_{p} = 0.85$ and $f_{p} = 2.0$ at particle densities
ranging from $n=0$ to $n=0.59$.
In the limit $n=0$, where the probe particle interacts only with 
the pinning, there is an initial pinned phase at low drive followed by a
transition to a moving state at a critical external drive $F^c_{ext}=1.7$. 
In the moving state the particle velocity is proportional to the
damping, $v \propto \eta$, representing Ohmic type behavior. 
For $n < 0.292$, the $v-F$ curves are similar
to the $n=0$ single particle limit, with only a slight
decrease in $v$ in the moving state for increasing $n$.  
For $n \geq 0.292$, 
in addition to the pinned regime at low drive and the 
Ohmic regime at high drive,
we find an intermediate non-Ohmic regime where the velocity 
undergoes pronounced fluctuations and is reduced below the Ohmic value.
Between the non-Ohmic and Ohmic regimes
there is a distinctive jump in the velocity 
at a drive $F_{ext}^t$
indicating a transition into a moving state with  
lower damping and reduced velocity fluctuations. 
The inset of Fig.~1 shows a blowup of the transition 
region for $n =  0.375$; the $n = 0.181$ curve is also presented for 
comparison. 
As $n$ increases, 
the velocity of the probe particle in the non-Ohmic regime decreases
and 
the transition drive $F_{ext}^t$ increases.
  
\begin{figure}
\includegraphics[width=3.5in]{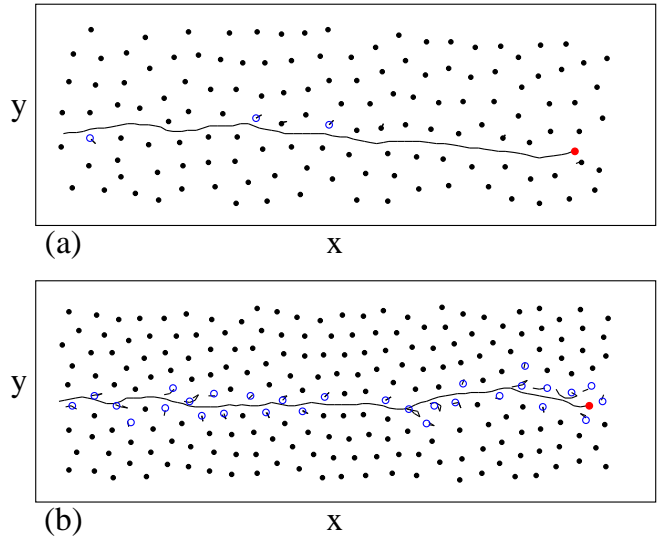}
\caption{Images of a $38 \times 12$ region of a sample with
$n_p=0.85$, $f_p=2.0$, and $F_{ext}=8.0$.
Large red dot: driven probe particle;
black dots: unperturbed surrounding
particles; 
open circles: surrounding particles perturbed by the passage of the probe
particle;
lines: particle trajectories. 
(a) The decoupled moving regime at $n=0.3$ has weak damping.
(b) The coupled moving regime at $n=0.5$ has high damping and significant
plastic distortions of the surrounding particles.
}
\end{figure}

To illustrate the origin of the different damping in the two moving states, 
in Fig.~2 we show the trajectories of the probe particle and the surrounding
particles in the weakly damped regime at $n=0.3$ 
and
in the strongly damped regime at $n=0.5$. 
For the weakly damped regime in Fig.~2(a), 
only small distortions of the
surrounding particles occur
as the probe 
particle moves through the sample. 
In contrast, Fig.~2(b) indicates that in the strongly damped regime, 
the surrounding particles are significantly disturbed by the passage of the
probe particle and undergo irreversible plastic distortion events.
In order for 
a plastic rearrangement 
to occur, 
a portion of the 
probe particle energy
must be transferred to the surrounding particles, 
increasing the
effective damping 
on the probe particle. 
The plastic distortions 
produce large fluctuations
in the probe particle velocity. 
In the weakly damped regime, the probe particle does not create
plastic distortions in the surrounding particles.
Thus, the transition we observe between the weakly and strongly damped
regimes occurs due to a viscous decoupling transition 
between the probe particle and the 
surrounding particles.

The ability of the probe particle to create plastic distortions 
is affected by the external drive, particle density,  
pinning strength, and pinning density. 
In general, a plastic rearrangement  
occurs
when a surrounding particle is both depinned and 
displaced by a distance $a_{0}/2$, where $a_{0}=n^{-1/2}$
is the average distance between particles.
The probe particle spends an average time $\delta t=a_0/F_{ext}$ 
interacting with a surrounding particle, and exerts a force
$F_d=-F_0\nabla V(a_0/2)$ on the particle.
In the absence of pinning, plastic distortions would occur when
$-F_0\nabla V(a_0/2)/F_{ext}> 1/2$,
and thus $F_{ext}^t \propto -\nabla V(1/(2\sqrt{n}))$.
The $v-F$ curves in Fig.~1 indicate that a transition between the 
high damping, plastic motion regime and the low damping, decoupled regime
occurs 
when 
the driving force 
$F_{ext}$ increases. 
When $F_{ext}$ is large, 
$F_{ext}\ge F_{ext}^t$, the probe particle passes the surrounding
particles so rapidly that it can only induce 
small displacements due to the short
interaction time $\delta t$. 
These displacements are too small to permit plastic rearrangements
to occur, and
the probe particle 
decouples from the surrounding particles.
For $F^c_{ext}<F_{ext}<F_{ext}^t$, plastic distortions occur and the probe
is in the high damping, coupled motion regime.
The 
lower limit of the coupled motion regime is determined by the fact that
the driving force $F_{ext}$ must exceed the critical
force $F^c_{ext}$ so that the probe particle itself 
remains depinned.

Figure 1 shows 
that the transition drive $F_{ext}^t$ increases with increasing
particle density $n$.  This is a result of the
increasing strength of the
interaction
$F_d$ between the probe particle and the surrounding particles
with increasing $n$.
At low 
$n$, 
the average distance
$a_o$
between particles
is large and 
$F_d$ 
is low, so the transition to decoupled motion $F_{ext}^t$ 
falls at low values of $F_{ext}$.
As $n$ increases, $F_d$ increases due to the decrease in $a_o$, 
and the plastic distortions persist up to higher values of $F_{ext}$,
corresponding to a higher value of $F_{ext}^t$.
In Fig.~3(a) we present a phase diagram 
for $F_{ext}$ versus $n$ taken from a series of $v-F$ curves.
The three phases, pinned (P), decoupled moving (DM), and coupled moving
(CM), are identified based on the values of $F_{ext}^c$ and
$F_{ext}^t$.
The range of driving forces over which coupled motion can occur increases
with increasing $n$.
For fixed drive, a velocity drop occurs when $n$ is increased above
the coupling-decoupling transition, as illustrated in the inset of Fig.~3(a)
for $F_{ext}=10$. 

\begin{figure}
\includegraphics[width=3.5in]{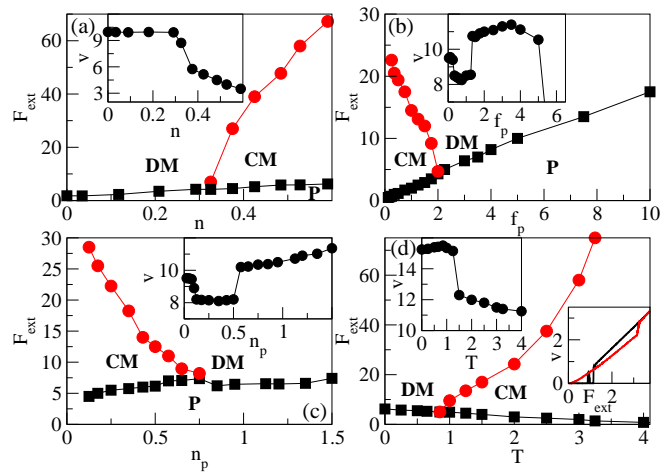}
\caption{(a) Phase diagram $F_{ext}$ versus $n$
at
$f_{p} = 2.0$ and $n_{p} = 0.85$ showing 
the pinned (P), coupled moving (CM), and decoupled moving (DM) 
regimes.
Squares: depinning threshold $F_{ext}^c$.  Circles: decoupling
threshold $F_{ext}^t$.
Inset: $v$ versus $n$ at $F_{ext} = 10.0$, showing a drop in $v$ above
the transition density of $n=0.3$.
(b) Phase diagram for $F_{ext}$ versus $f_{p}$ 
at $n_{p} = 0.85$ and $n = 0.292$.
Inset: $v$ versus $f_{p}$ at $F_{ext} = 12.0$ 
shows that $v$ can increase with increasing $f_{p}$. 
(c) Phase diagram for $F_{ext}$ versus $n_{p}$ 
for $f_{p} = 3.0$ and $n = 0.292$. 
Inset: $v$ versus $n_{p}$ for $F_{ext} = 12.0$. (d) 
Phase diagram for $F_{ext}$ versus $T$ with $n_{p} = 0.85$, $f_{p} = 3.0$, 
and $n = 0.292$.
Left inset: $v$ vs T for $F_{ext} = 16.0$. 
Right inset: $v$ vs $F_{ext}$ for 
superconducting
vortices with $n=0.44$ and $n_p=0.46$.
Lower right curve: $f_{p} = 0.25$; upper right curve: $f_{p} = 1.5$. 
}
\end{figure}

To determine the influence of the pinning on the coupling-decoupling
transition, we vary the pinning force $f_p$ in a system with fixed $n = 0.292$
and $n_{p}= 0.85$, and plot the resulting phase diagram in Fig.~3(b).
The depinning force increases linearly with $f_p$, $F_{ext}^c \propto f_p$. 
For very low $f_{p}$ the system  
responds as if there were no pinning, a coupled motion
regime which was previously 
explored in Ref.~\cite{Hastings}.
At low $f_{p}$, the probe particle is strongly coupled
to the surrounding particles and generates significant plastic distortions.
Since the pinning force tends to localize the particles and therefore
competes with $F_d$ in determining whether the
surrounding particles can undergo plastic motion, we expect plastic
distortions to occur in the presence of pinning when
$(F_d-f_p)\delta t>a_0/2$, giving $F_{ext}^t\propto (F_d-f_p)$.
This is in agreement with Fig.~3(b), where $F^t_{ext}$ decreases 
linearly with $f_p$. The
coupled motion disappears 
for $f_p>2.0$
when the probe particle can no longer depin the surrounding particles.

For fixed drive, the probe particle velocity $v$ initially decreases with
increasing $f_p$, as shown in the inset of Fig.~3(b) for $F_{ext}=12.0$.
The pinning increases the effective viscosity in the coupled moving regime
compared to the $f_p=0$ case by
increasing 
the number of irreversible events that occur. 
As $f_p$ increases, a surrounding particle that has been depinned and displaced 
by the probe particle can more easily be repinned at a new pinning site, rather
than returning to its previous position in a reversible event.
When $f_{p}$ is further increased, a sharp increase in $v$ occurs at 
the decoupling transition where  
the plastic distortions are lost. 
We note that $v$ in the decoupled regime is {\it higher} than 
the value of $v$ at $f_{p} = 0$.
These results show that increasing the strength of the pinning 
can cause a counterintuitive {\it increase} in the
velocity (or decrease in the damping) of the probe particle.
At large enough
$f_{p}$, the probe particle itself becomes pinned. 

In Fig.~3(c) we show the phase diagram for $F_{ext}$ versus 
pinning density $n_{p}$ 
for a system with $f_{p} = 3.0$ and $n = 0.292$. 
There is little change in $F_{ext}^c$ with $n_p$, while $F_{ext}^t$ decreases
with $n_p$ until the coupled motion regime disappears for $n_p>0.75$.
When $n_p$ increases, the average spacing 
between adjacent pinning sites decreases, and 
a surrounding particle that has been 
depinned by the probe particle can be trapped by a new pinning site before it
has moved far enough to allow a plastic rearrangement 
to occur.
The inset of Fig.~3(c) illustrates the velocity of the 
probe particle versus $n_{p}$ at fixed $F_{ext} = 12.0$. 
For very low $n_{p}$ the behavior is similar to the pin-free case.
The effective viscosity increases sharply at $n_{p} \approx 0.125$ when the 
system enters the strongly damped, coupled motion regime. 
For $n_{p}>0.5$, 
the viscosity drops sharply at the onset of the decoupled
motion regime, and then monotonically decreases 
for increasing $n_{p}$.
Increasing the pinning density 
can counterintuitively {\it increase} 
the velocity of the probe particle 
$v$ above the value
at $n_{p} = 0$.   

The phase diagram for $F_{ext}$ vs temperature $T$ is shown in Fig.~3(d)
for a system with $f_{p} = 3.0$, $n_{p} = 0.85$, and $n = 0.292$.  
At $T = 0$ the probe particle undergoes decoupled motion 
for $F_{ext}>F_{ext}^c$. 
As $T$ increases, the effectiveness of the pinning
is reduced, $F_{ext}^c$ decreases, and the probe 
particle recouples to the surrounding 
particles.  This produces a velocity drop with increasing $T$, as 
shown in the left inset of Fig.~3(d) for $F_{ext}=16.0$.
A transition back to the decoupled moving regime occurs
for $F_{ext}>F_{ext}^t$. 
The overall structure of the phase diagram in Fig.~3(d) 
is very similar to the dynamical phase diagram 
for vortices in type-II superconductors 
moving over random disorder, where a transition from  
plastic to elastic flow occurs as the driving force on the
vortex lattice is increased 
\cite{Koshelev}. Our results 
indicate that a similar effect can occur even 
when only a single particle is driven.

We have performed similar simulations 
for vortices in type-II superconductors. 
In the right inset of Fig.~3 we plot the $v-F$ curves 
from a vortex sample with $n_{p} = 0.46$, 
$n_{v} = 0.44$, and $r_p = 0.2$ in the strong pinning regime with 
$f_p=1.5$ and in the weak pinning regime with $f_p=0.25$.
The $v-F$ curves have the same trend 
seen in Fig.~1 and Fig.~3(b).  The $f_p=0.25$
sample exhibits 
coupled motion with high damping, while the 
$f_p=1.5$ 
sample is in the low damping, decoupled motion regime.
A decoupling transition 
occurs with increasing driving force 
for the weakly pinned sample.
This suggests that our results should be generic to single driven probe 
particles moving through a background of repulsively interacting particles
in the presence of quenched disorder.

In summary, we show that when a single probe particle is driven 
through an assembly of other particles
in the presence of quenched disorder, a novel viscous decoupling 
transition can occur 
between the probe particle and the surrounding particles
which is controlled by the strength and density of the quenched disorder. 
This transition is from a highly damped regime, where the probe particle 
depins the surrounding particles and produces irreversible plastic 
distortions,
to a low damping state, where the probe particle does not couple to the 
surrounding particles.
Increasing the pinning strength or density reduces the coupling between the
probe and background particles, producing the counterintuitive result that
increasing the strength or density of the quenched disorder
{\it increases} the velocity of the probe particle.  
The decoupling transition appears as a clear signal in the effective damping on the
probe particle and produces a distinct feature in the velocity force curve.
In addition to colloids with Yukawa interactions, our results 
should be general to other systems of interacting particles with quenched
disorder,
including vortices in type-II superconductors. 
 
This work was carried out under the auspices of the 
NNSA of the 
U.S. DoE
at 
LANL
under Contract No.
DE-AC52-06NA25396.

\end{document}